\documentstyle[prb,aps]{revtex}
\begin{document}
\draft
\title{Stochastic model of hysteresis}
\author{L. P\'al}
\address{KFKI Atomic Energy Research Institute
1525 Budapest, P.O.B. 49. Hungary}
\date{\today}
 
\maketitle
\begin{abstract}
The methods of the probability theory have been used in order to
build up a new model of hysteresis which is different from the
well-known  Preisach model. It is assumed that the system
consists of large number of abstract particles in which the
variation of an external control parameter (e. g., the magnetic
field) may result in transitions between two states
${\cal S}^{(+)}$ and ${\cal S}^{(-)}$. The state of a particle
is characterized by the value $+1$ or $-1$ of a random variable
(e. g., the magnetization direction parallel or antiparallel
to the magnetic field). The transitions are  governed by
two further random variables corresponding to the ${\cal S}^{(-)}
\Rightarrow {\cal S}^{(+)}$ and the ${\cal S}^{(+)}
\Rightarrow {\cal S}^{(-)}$ transitions (e. g., "up switching"
and "down switching magnetic field").
The method presented here makes possible
to calculate the probability distribution and consequently the
expectation value of the number of particles in the
${\cal S}^{(+)}$ (or ${\cal S}^{(-)}$) state for both increasing
and decreasing parameter values, i. e., the hysteresis curves of
the transitions can be determined. It turns out that
the reversal points of the
control parameter are Markov points which determine the
stochastic evolution of the process. It has been shown that
the branches of the hysteresis loop are converging to fixed
limit curves
when the number of cyclic back-and-forth variations of
the control parameter between two consecutive reversal
points is large enough. This convergence to limit curves gives a
clear explanation of the accommodation process. The accommodated
minor loops show the return-point memory property but this
property is obviously absent in the case of
non-accommodated minor loops which
are not congruent and generally not closed.
In contrast to the traditional Preisach model the reversal point
susceptibilities are non-zero finite values.
The stochastic model can provide a surprisingly good
approximation of the Raylaigh quadratic law when the external
parameter varies between two sufficiently small values.
The practical benefits of the model can be seen in the
numerical analysis of the derived equations. On one hand
the calculated curves are in good qualitative agreement with
the experimental observations and on the other hand,
the estimation of the joint distribution
function of the up and down switching fields can be performed by
using the measured hysteresis curves.
\end{abstract}
 
\vspace{1cm}
 
\section{Introduction}
 
The phenomenon of the hysteresis, understood in a general sense,
has been investigated so intensively for many decades that
any list of references would be far from complete by any
standards. It is very fortunate that in the last few years
outstanding monographs \cite{mayergoyz91}$^{,}$
\cite{visintin94}$^,$ \cite{ivanyi97}$^,$ \cite{bertotti98}
have been published in this field, and thus the author does
not feel obliged to cite the large amount of old but
important references. However, it is considered important to
mention two papers of G. K\'ad\'ar
\cite{kadar87}$^{,}$\cite{kadar89}
whose work has played a stimulating role in getting to the idea
of reconsideration of the hysteresis theory by the present author.
There is no doubt that the abstract reformulation of the Preisach
model \cite{preisach35} given by Krasnoselskii and A. Pokrovskii
\cite{krasnoselskii71} and summarized by I. D. Mayergoyz
\cite{mayergoyz91} in his book resulted in an improved
mathematical clarity in the hysteresis theory,
but the stochastic nature of the hysteresis
still has not been treated with sufficient mathematical rigor.
 
The aim of the present paper is to define a stochastic model of
hysteresis and to derive exact equations for the probability
distribution functions describing the state variations in
hysteretic processes as a function of increasing as well as
decreasing control parameters. The vocabulary of magnetic
hysteresis will be used for convenience from now on, however, the
concepts can easily be generalized for any hysteretic phenomenon.
 
\section{Description of the model}
 
Let us assume that the unit volume of the system consists of
many small abstract regions, called "particles" which are
characterized by four random variables: $\mu, \lambda, \chi_{d}$
and $\chi_{u}$. The absolute value of the particle magnetization
is denoted by $\mu$. If the particle magnetization is parallel
(antiparallel) to the external magnetic field $H$ then the
particle is in the state ${\cal S}^{(+)}$ $({\cal S}^{(-)})$ and
$\lambda = 1 (-1)$. The random value $\chi_{d}$ corresponds to a
local field at which the state ${\cal S}^{(+)}$  jumps to the
state ${\cal S}^{(-)}$ and similarly the
${\cal S}^{(-)} \Rightarrow {\cal S}^{(+)}$ transition occurs
at the random  local field $\chi_{u}$. For simplicity
the $\chi_{u}$ and $\chi_{d}$ will be called as U-field and
D-field. These two random variables express the obvious fact
that each particle "feels" not only the external magnetic field
$H$, but also the interaction field due
to the adjacent particles and the
random field originated from the inhomogenities of the
surrounding environment. These particles characterized by the
random variables $\mu, \lambda, \chi_{d}$ and $\chi_{u}$
can be regarded as "independent" abstract elements of
the system, and they will be called {\em "hysterons"}.
 
Figure \ref{fig1} illustrates a possible realization of
transitions ${\cal S}^{(+)} \Leftrightarrow {\cal S}^{(-)}$ of a
hysteron. The transition curves form a random rectangular
hysteresis loop which is almost in all cases asymmetrical in the
co-ordinate system of magnetization versus external magnetic field
since the U- and D-fields are supposed to be random.
 
Let us denote by $h_{1}, h_{2}, \ldots, h_{N}$ the hysterons in a
system, and let ${\cal N}^{(+)}$ be the set of indices and
$n^{(+)}$ the number of hysterons
which are in the state ${\cal S}^{(+)}$
at a given external field $H$. In this case the magnetization
of the system is given by the stochastic equation
\[ \delta_{n^{(+)}} = \sum_{k=1}^{N} \lambda_{k}\;\mu_{k}, \]
where $\mu_{k}$ is the absolute value of the magnetization of
the hysteron $h_{k}$, while
\[ \lambda_{k} = \left\{ \begin{array}{ll}
           1, & \mbox{if $k \in {\cal N}^{(+)}$,} \\
               \mbox{} & \mbox{} \\
          - 1, & \mbox{if $k  \ni {\cal N}^{(+)}$.}
 \end{array} \right. \]
Since the random variables $\mu_{1}, \mu_{2}, \ldots, \mu_{N}$
are mutually independent and have the same probability
distribution function
\[ {\bf P}\{\mu_{k} \leq x \} = L(x), \;\;\;\;\; \forall \;\;
k=1, 2, \ldots, N, \]
it is obvious that the characteristic function of the
distribution function
\begin{equation}
{\bf P}\{\delta_{n^{(+)}} \leq x \} = R_{n^{(+)}}(x)
\label{eq1}
\end{equation}
can be written in the form
\begin{equation}
\Phi_{n^{(+)}}(\omega) = \int_{-\infty}^{+\infty}
e^{i\omega x}\; dR_{n^{(+)}}(x) =
[\varphi(-\omega)]^{N}\;\left[\frac{\varphi(\omega)}
{\varphi(-\omega)}\right]^{n^{(+)}},
\label{eq2}
\end{equation}
where
\begin{equation}
\varphi(\omega) = \int_{-\infty}^{+\infty} e^{i\omega x}\;
dL(x) = \int_{0}^{+\infty} e^{i\omega x}\;dL(x).
\label{eq3}
\end{equation}
In order to calculate the characteristic function
\begin{equation}
\Phi(H, \omega) = \sum_{n^{(+)}=0}^{N}
\Phi_{n^{(+)}}(\omega)\;p_{n^{(+)}}(H)  =
[\varphi(-\omega)]^{N}\;\sum_{n^{(+)}=0}^{N}
p_{n^{(+)}}(H)\;\left[\frac{\varphi(\omega)}
{\varphi(-\omega)}\right]^{n^{(+)}},
\label{eq4}
\end{equation}
we need the probability of finding
$n^{(+)}$ hysterons in the state ${\cal S}^{(+)}$ at the external
field $H$ which is the endpoint of a well-defined
magnetization prehistory. The determination of this probability
and the derivation of the equations for "up" and "down"
magnetizations versus magnetic field will be the task of the next
section.
 
\section{Derivation of the fundamental equations}
 
\subsection{Some basic relations}
 
Let us denote by $H(x, y|{\cal C})$ the joint distribution
function of the random U- and D-fields.
From the physical point of view it is quite obvious that the
U-field cannot be smaller than the D-field, so the
stochastic inequality $\chi_{u} \geq \chi_{d}$ must be
satisfied. It is easy to show \cite{pal95} that the joint
distribution function of $\chi_{u}$ and $\chi_{d}$ satisfying the
condition ${\cal C} = \{\chi_{u} \geq \chi_{d}\}$ can be written
in the form
\begin{equation}
{\bf P}\{\chi_{u} \leq x, \chi_{d} \leq y| \chi_{u} \geq \chi_{d}
\} = H(x, y|{\cal C}) =
\frac{\int_{-\infty}^{x} dx' \int_{-\infty}^{y}
h(x', y')\; \Delta(x'-y')\;dy'}
{\int_{-\infty}^{+\infty} dx' \int_{-\infty}^{x'}
h(x', y')\;dy'},
\label{eq5}
\end{equation}
where $\Delta(x)$ is the unit step function. It is clear that
the joint density function of the U- and D-fields can be given by
\begin{equation}
h(x, y|{\cal C}) =
\frac{h(x, y)}{\int_{-\infty}^{+\infty} dx' \int_{-\infty}^{x'}
h(x', y')\;dy'}\;\Delta(x-y),
\label{eq6}
\end{equation}
provided that the condition ${\cal C}$ is valid.
 
We need in the sequel two conditional probability distribution
functions:
\begin{equation}
{\bf P}\{ \chi_{u} \leq x|{\cal C}\} = H(x, \infty|{\cal C}) =
F_{u}(x|{\cal C}) = \frac{\int_{-\infty}^{x}
dx'\;\int_{-\infty}^{x'} h(x', y')\;dy'}
{\int_{-\infty}^{+\infty} dx' \int_{-\infty}^{x'}
h(x', y')\;dy'}
\label{eq7}
\end{equation}
and
\begin{equation}
{\bf P}\{ \chi_{d} \leq y|{\cal C}\} = H(\infty, y|{\cal C}) =
F_{d}(y|{\cal C}) = \frac{\int_{-\infty}^{y}
dy'\;\int_{y'}^{+\infty} h(x', y')\;dx'}
{\int_{-\infty}^{+\infty} dy' \int_{y'}^{+\infty}
h(x', y')\;dx'}.
\label{eq8}
\end{equation}
Evidently $F_{u}(x|{\cal C})$ is the probability that the
U-field of a given hysteron is not larger than $x$, while
$F_{d}(y|{\cal C})$ is the probability that the D-field is not
larger than $y$ assuming in both cases that the condition ${\cal
C}$ is fulfilled. By using the Dirichlet's theorem for changing
the sequence of integration it is obvious that
\[\int_{-\infty}^{+\infty} dx' \int_{-\infty}^{x'}
h(x', y')\;dy' =
\int_{-\infty}^{+\infty} dy' \int_{y'}^{+\infty}
h(x', y')\;dx'. \]
 
For the sake of further considerations it is necesarry to
introduce two {\em transition probabilities} denoted
by $w_{u}(H_{l}\!\uparrow\!H)$
and $w_{d}(H_{u}\!\downarrow\!H)$. Let $H_{l}$ be a fixed value
of the external magnetic field and let us suppose that the state
of a given hysteron is ${\cal S}^{(-)}$ at $H_{l}$. By using
elementary theorems, it can be proved that if the external
field increases monotonically from $H_{l}$ to $H \geq H_{l}$ then
\begin{equation}
w_{u}(H_{l}\!\uparrow\!H) = \frac{\int_{H_{l}}^{H} dx\;
\int_{-\infty}^{x} h(x, y)\;dy}{\int_{H_{l}}^{+\infty} dx\;
\int_{-\infty}^{x} h(x, y)\;dy}
\label{eq9}
\end{equation}
is the probability of the transition ${\cal S}^{(-)}
\Rightarrow {\cal S}^{(+)}$ occuring in the interval
$[H_{l}\!\uparrow\!H]$. Similarly, let $H_{u}$ be an other fixed
value and ${\cal S}^{(+)}$ the state of a hysteron at $H_{u}$. If
the external field decreases now monotonically from $H_{u}$ to $H
\leq H_{u}$ then
\begin{equation}
w_{d}(H_{u}\!\downarrow\!H) = \frac{\int_{H}^{H_{u}} dy\;
\int_{y}^{+\infty} h(x, y)\;dx}{\int_{-\infty}^{H_{u}} dy\;
\int_{y}^{+\infty} h(x, y)\;dx}
\label{eq10}
\end{equation}
gives the probability of the transition ${\cal S}^{(+)}
\Rightarrow {\cal S}^{(-)}$ occuring in the interval
$[H_{u}\!\downarrow\!H]$.
 
\subsection{Stochastic magnetizing process}
 
Let us introduce the "time parameter" $t \in [0, +\infty]$ and
define a real valued, external field function
$H(t)$ which consists of monotone increasing and decreasing
sections of different length. Denote by $H_{1}, H_{2}, \ldots,
H_{j}, \ldots,$ the extremum values of the function $H(t)$
belonging to the subsequent time points
$t_{1} < t_{2} < \cdots < t_{j} < \cdots$.
It is clear that if $H(t_{j}) = H_{j}$ is
a local maximum then $H(t_{j-1}) = H_{j-1}$ and $H(t_{j+1}) =
H_{j+1}$ must be local minimums which are not necessarily equal.
In the following the  sequence $\{H_{j}\}$  will be called
{\em magnetizing path}  and the elements of this sequence
are called {\em points of reversal}.
If the functions $H^{(1)}(t)$ and $H^{(2)}(t)$
have the same magnetizing path then they are said to be
equivalent for any magnetizing process irrespective of the form
of the time function  between the individual extrema. In Fig.
\ref{fig2} two {\em equivalent $H(t)$ functions} are seen.
The sequence of extrema is the same for both curves, but the
time distance and the shape of sections between the consecutive
extrema are different.
 
It is assumed that the magnetizing process which consists
of random transitions ${\cal S}^{(+)} \Leftrightarrow
{\cal S}^{(-)}$ of hysterons does not "feel" the variation speed
of $H(t)$ between the consecutive extremum values, i. e., the
magnetizing process is {\em static}.
The evolution of the process in each subinterval
$[H_{j}, H_{j+1}], \;\; j=1,2,\ldots,$
is stochasically determined by the extremum $H_{j}$
and by the actual values of $H(t)$ following $H_{j}$, but the
process does not depend on the time derivative of
$H(t)$. This property is called {\em rate independence} in the
non-stochastic theory of hysteresis \cite{visintin94} but it will
be applied in this stochastic theory too.
Denote the maximum reversal fields by odd  and the minimum
ones by even indices. In this case it is clear that
\[ H_{2k-1} \geq H_{2k} \leq H_{2k+1}, \]
and naturally any one of the inequalities $H_{2k+1} \geq H_{2k-1}$
and $H_{2k+1} \leq H_{2k-1}$ can be valid.
 
Now, let us define {\em the random function}
$\xi_{d}^{(+)}(H_{2k-1}\!\downarrow\!H)$
which gives the number of hysterons in the state
${\cal S}^{(+)}$ at the decreasing
external field $H$ belonging to the interval
$[H_{2k-1}\!\downarrow\!H_{2k}]$. Similarly, denote by
$\xi_{u}^{(+)}(H_{2k}\!\uparrow\!H)$ the number of hysterons in
the state ${\cal S}^{(+)}$ at the increasing external field $H$
belonging to the interval $[H_{2k}\!\uparrow\!H_{2k+1}]$.
 
We suppose that at the starting point of the magnetizing process
each hysteron is in the ${\cal S}^{(-)}$ state, that is
the system is in the state of negative saturation.
In the following this fact will be expressed by
the stochastic equation $\xi_{start}^{(+)} = 0$.
Evidently any other state of the system could as well be chosen
for the starting point, this choice, however, does not really
matter since the influence of the starting state
on the evolution of the process
--- as it will be shown --- disappears very rapidly.
 
In order to describe the magnetizing process we should
determine two probabilities. One of them is
\begin{equation}
{\bf P}\{\xi_{d}^{(+)}(H_{2k+1}\!\downarrow\!H) =
n_{2k+1}^{(d)}(H)|
\xi_{start}^{(+)}=0\} = p_{2k+1}^{(d)}[H_{2k+1}\!\downarrow\!H,
n_{2k+1}^{(d)}(H)|0],
\label{eq11}
\end{equation}
and the other is
\begin{equation}
{\bf P}\{\xi_{u}^{(+)}(H_{2k}\!\uparrow\!H) = n_{2k}^{(u)}(H)|
\xi_{start}^{(+)}=0\} = p_{2k}^{(u)}[H_{2k}\!\uparrow\!H,
n_{2k}^{(u)}(H)|0].
\label{eq12}
\end{equation}
It is important to note that the reversal points (extremum values)
\[ H_{1}, H_{2}, \ldots, H_{2k-1}, H_{2k}, H_{2k+1}, \ldots, \]
are {\em Markov-points of the stochastic processes}
$\xi_{u}^{(+)}(H_{j}\!\uparrow\!H)$
and
$\xi_{d}^{(+)}(H_{j+1}\!\downarrow\!H)$,
and therefore we can write the following equations:
\[ p_{2k+1}^{(d)}[H_{2k+1}\!\downarrow\!H, n_{2k+1}^{(d)}(H)|0] =
\sum_{n_{2k}^{(u)}(H_{2k+1})=0}^{N}
p_{2k+1}^{(d)}[H_{2k+1}\!\downarrow\!H, n_{2k+1}^{(d)}(H)|
n_{2k}^{(u)}(H_{2k+1})]\times \]
\begin{equation}
\times
p_{2k}^{(u)}[H_{2k}\!\uparrow\!H_{2k+1},n_{2k}^{(u)}(H_{2k+1})|0],
\label{eq13}
\end{equation}
and
\[ p_{2k}^{(u)}[H_{2k}\!\uparrow\!H, n_{2k}^{(u)}(H)|0] =
\sum_{n_{2k-1}^{(d)}(H_{2k})=0}^{N}
p_{2k}^{(u)}[H_{2k}\!\uparrow\!H, n_{2k}^{(u)}(H)|
n_{2k-1}^{(d)}(H_{2k})]\times \]
\begin{equation}
\times p_{2k-1}^{(d)}[H_{2k-1}\!\downarrow\!H_{2k},
n_{2k-1}^{(d)}(H_{2k})|0].
\label{eq14}
\end{equation}
 
As it has already been mentioned the hysterons can be
regarded as independent of each other particles and, therefore,
it is an easy task to determine the
probability that the number of ${\cal S}^{(+)}$-hysterons
is exactly equal to a  non-negative integer not larger than $N$,
at an either decreasing or increasing  external field $H$
provided that the number of ${\cal S}^{(+)}$-hysterons is known
at the last reversal point before arriving at $H$.
 
The probability
$p_{2k+1}^{(d)}[H_{2k+1}\!\downarrow\!H, n_{2k+1}^{(d)}(H)|
n_{2k}^{(u)}(H_{2k+1})]$ can be obtained as a result of the
following consideration.  If the number of  ${\cal
S}^{(+)}$-hysterons at the reversal point $H_{2k+1}$ is equal to
$n_{2k}^{(u)}(H_{2k+1})$, then --- in order to have
$n_{2k+1}^{(d)}(H)$ hysterons in the state ${\cal S}^{(+)}$ at
the external field $H \leq H_{2k+1}$ --- exactly
$n_{2k}^{(u)}(H_{2k+1}) - n_{2k+1}^{(d)}(H)$ hysterons of state
${\cal S}^{(+)}$ have to transform to the state ${\cal
S}^{(-)}$ in the interval $[H_{2k+1}\!\downarrow\!H]$.
It is obvious that the probability of this event can be given by
\[ p_{2k+1}^{(d)}[H_{2k+1}\!\downarrow\!H, n_{2k+1}^{(d)}(H)|
n_{2k}^{(u)}(H_{2k+1})] = \left(\begin{array}{c}
n_{2k}^{(u)}(H_{2k+1}) \\ n_{2k+1}^{(d)}(H) \end{array} \right)
\times \]
\begin{equation}
\times \left[w_{d}(H_{2k+1}\!\downarrow\!H)\right]^
{n_{2k}^{(u)}(H_{2k+1}) - n_{2k+1}^{(d)}(H)}\;
\left[1 - w_{d}(H_{2k+1}\!\downarrow\!H)\right]^
{n_{2k+1}^{(d)}(H)}.
\label{eq15}
\end{equation}
Similarly, to determine the probability
$p_{2k}^{(u)}[H_{2k}\!\uparrow\!H, n_{2k}^{(u)}(H)|
n_{2k-1}^{(d)}(H_{2k})]$ one has to recognize that if
the number of ${\cal S}^{(-)}$-hysterons
at the reversal point $H_{2k}$ is equal to
$N - n_{2k-1}^{(d)}(H_{2k})$, then --- in order to have
$n_{2k}^{(u)}(H)$ hysterons in the state ${\cal S}^{(+)}$  at the
external field $H \geq H_{2k}$ --- exactly
$n_{2k}^{(u)}(H) - n_{2k-1}^{(d)}(H_{2k-1})$ hysterons of state
${\cal S}^{(-)}$ have to transform to the state
${\cal S}^{(+)}$ in the interval
$[H_{2k}\!\uparrow\!H]$. The probability of this event is given by
\[ p_{2k}^{(u)}[H_{2k}\!\uparrow\!H, n_{2k}^{(u)}(H)|
n_{2k-1}^{(d)}(H_{2k})] = \left(\begin{array}{c}
N-n_{2k-1}^{(d)}(H_{2k}) \\ n_{2k}^{(u)}(H)-
n_{2k-1}^{(d)}(H_{2k})  \end{array} \right) \times \]
\begin{equation}
\times \left[w_{u}(H_{2k}\!\uparrow\!H)\right]^
{n_{2k}^{(u)}(H) - n_{2k-1}^{(d)}(H_{2k})}\;
\left[1 - w_{u}(H_{2k}\!\uparrow\!H)\right]^
{N - n_{2k}^{(u)}(H)}.
\label{eq16}
\end{equation}
 
In order to symplify the further calculations let us introduce
{\em the generating func\-tions}
\begin{equation}
\Gamma_{2k+1}^{(d)}(H_{2k+1}\!\downarrow\!H, z) =
\sum_{n_{2k+1}^{(d)}(H)=0}^{N}
p_{2k+1}^{(d)}[H_{2k+1}\!\downarrow\!H, n_{2k+1}^{(d)}(H)|0]\;
z^{n_{2k+1}^{(d)}(H)}
\label{eq17}
\end{equation}
and
\begin{equation}
\Gamma_{2k}^{(u)}(H_{2k}\!\uparrow\!H, z) =
\sum_{n_{2k}^{(u)}(H)=0}^{N}
p_{2k}^{(u)}[H_{2k}\!\uparrow\!H, n_{2k}^{(u)}(H)|0]\;
z^{n_{2k}^{(u)}(H)}.
\label{eq18}
\end{equation}
By using the Eqs. (\ref{eq13}) and (\ref{eq15}) we get
{\em the first fundamental equation} in the form
\begin{equation}
\Gamma_{2k+1}^{(d)}(H_{2k+1}\!\downarrow\!H, z) =
\Gamma_{2k}^{(u)}[H_{2k}\!\uparrow\!H_{2k+1}, a(H_{2k+1}, H, z)],
\label{eq19}
\end{equation}
where
\begin{equation}
a(H_{2k+1}, H, z) = w_{d}(H_{2k+1}\!\downarrow\!H) +
[1 - w_{d}(H_{2k+1}\!\downarrow\!H)]\;z.
\label{eq20}
\end{equation}
{\em The second fundamental equation} follows from the relations
(\ref{eq14}) and (\ref{eq16}). We have
\begin{equation}
\Gamma_{2k}^{(u)}(H_{2k}\!\uparrow\!H, z) =
[c(H_{2k}, H, z)]^{N}\;\Gamma_{2k-1}^{(d)}
[H_{2k-1}\!\downarrow\!H_{2k}, b(H_{2k}, H, z)],
\label{eq21}
\end{equation}
where
\begin{equation}
c(H_{2k}, H, z) = 1 - (1 - z)\;w_{u}(H_{2k}\!\uparrow\!H),
\label{eq22}
\end{equation}
and
\begin{equation}
b(H_{2k}, H, z) = \frac{z}{c(H_{2k}, H, z)}.
\label{eq23}
\end{equation}
 
Now we will derive the characteristic function of the
probability that the system magnetization is not larger than
$x$ at a decreasing  external field $H$ which follows the last
reversal point $H_{2k+1}$. Introducing the notation
\begin{equation}
\psi(\omega) = \frac{\varphi(\omega)}{\varphi(-\omega)},
\label{eq24}
\end{equation}
and by using the relation (\ref{eq17}) we obtain from
Eq. (\ref{eq4})
\begin{equation}
\Phi_{d}(H_{2k+1}\!\downarrow\!H, \omega) =
[\varphi(-\omega)]^{N}\;\Gamma_{2k+1}^{(d)}
[H_{2k+1}\!\downarrow\!H, \psi(\omega)].
\label{eq25}
\end{equation}
Similarly, if we use Eq. (\ref{eq18}) then
the characteristic function of the probability that the
magnetization of the system is not larger than $x$ at an
increasing external field $H$ after the last reversal point
$H_{2k}$ can be obtained from Eq. (\ref{eq4})
in the form
\begin{equation}
\Phi_{u}(H_{2k}\!\uparrow\!H, \omega) =
[\varphi(-\omega)]^{N}\;\Gamma_{2k}^{(u)}
[H_{2k}\!\uparrow\!H, \psi(\omega)].
\label{eq26}
\end{equation}
These two characteristic functions describe completely the
stochastic behaviour of the magnetizing process in both
increasing and decreasing external magnetic fields. It is apparent
from the above considerations that the stochastic model
developed by us has been built up  without any reference to a
particular nature of hysteresis and therefore, its generality
is at least as high as that of the Krasnoselskii and Pokrovskii
\cite{krasnoselskii71} model.
 
\subsection{Calculation of the hysteresis curves}
 
The expectation value of the magnetic moment $\mu_{k}$ due to the
$k$-th hysteron can be given by
\[ {\bf E}\{\mu_{k}\} = i^{-1}\;\left[\frac{d\varphi(\omega)}
{d\omega}\right]_{\omega=0} = M_{s}. \]
By using this expression we can write the expectation value of the
magnetization of the system at the decreasing external field $H$
following the reversal point $H_{2k+1}$ in the form
\[ i^{-1}\;\left[\frac{d\Phi_{d}(H_{2k+1}\!\downarrow\!H, \omega)}
{d\omega}\right]_{\omega=0} = M_{2k+1}^{(d)}
(H_{2k+1}\!\downarrow\!H|0) =  \]
\begin{equation}
= 2M_{s}\;N_{2k+1}^{(d)}(H_{2k+1}\!\downarrow\!H|0) -
N\;M_{s},
\label{eq27}
\end{equation}
where
\begin{equation}
N_{2k+1}^{(d)}(H_{2k+1}\!\downarrow\!H|0) =
\left[\frac{d\Gamma_{2k+1}^{(d)}(H_{2k+1}\!\downarrow\!H, z)}
{dz}\right]_{z=1}.
\label{eq28}
\end{equation}
From the fundamental Eq. (\ref{eq19}) we obtain
\begin{equation}
N_{2k+1}^{(d)}(H_{2k+1}\!\downarrow\!H|0) =
N_{2k}^{(u)}(H_{2k}\!\uparrow\!H_{2k+1}|0)\;[1 -
w_{d}(H_{2k+1}\!\downarrow\!H)].
\label{eq29}
\end{equation}
The expectation value of the magnetization of the
system at the external magnetic field $H$ increasing
after the reversal field $H_{2k}$ can be obtained from the
equation
\[ i^{-1}\;\left[\frac{d\Phi_{u}(H_{2k}\!\uparrow\!H, \omega)}
{d\omega}\right]_{\omega=0} = M_{2k}^{(u)}
(H_{2k}\!\uparrow\!H|0) =  \]
\begin{equation}
= 2M_{s}\;N_{2k}^{(u)}(H_{2k}\!\uparrow\!H|0) -
N\;M_{s},
\label{eq30}
\end{equation}
where
\begin{equation}
N_{2k}^{(u)}(H_{2k}\!\uparrow\!H|0) =
\left[\frac{d\Gamma_{2k}^{(u)}(H_{2k}\!\uparrow\!H, z)}
{dz}\right]_{z=1}.
\label{eq31}
\end{equation}
The recursive relation
\begin{equation}
N_{2k}^{(u)}(H_{2k}\!\uparrow\!H|0) =
N\;w_{u}(H_{2k}\!\uparrow\!H) +
N_{2k-1}^{(d)}(H_{2k-1}\!\downarrow\!H_{2k}|0)\;
[1 - w_{u}(H_{2k}\!\uparrow\!H)]
\label{eq32}
\end{equation}
follows from the other fundamental Eq. (\ref{eq21}).
 
By intoducing {\em the relative magnetizations}
\begin{equation}
m_{2k+1}^{(d)}(H_{2k+1}\!\downarrow\!H|0) =
\frac{1}{N\;M_{s}}\;M_{2k+1}^{(d)}(H_{2k+1}\!\downarrow\!H|0),
\label{eq33}
\end{equation}
and
\begin{equation}
m_{2k}^{(u)}(H_{2k}\!\uparrow\!H|0) =
\frac{1}{N\;M_{s}}\;M_{2k}^{(u)}(H_{2k}\!\uparrow\!H|0),
\label{eq34}
\end{equation}
from Eqs. (\ref{eq27}), (\ref{eq29}) and (\ref{eq30}),
(\ref{eq32}) after elementary calculations the following recursive
relations are obtained:
\[ m_{2k+1}^{(d)}(H_{2k+1}\!\downarrow\!H|0) = \]
\begin{equation}
= \left[1 + m_{2k}^{(u)}(H_{2k}\!\uparrow\!H_{2k+1}|0)\right]\;
\left[1 - w_{d}(H_{2k+1}\!\downarrow\!H)\right] - 1,
\label{eq35}
\end{equation}
and
\[ m_{2k}^{(u)}(H_{2k}\!\uparrow\!H|0) = \]
\begin{equation}
= 2\;w_{u}(H_{2k}\!\uparrow\!H) +
\left[1 +
m_{2k-1}^{(d)}(H_{2k-1}\!\downarrow\!H_{2k}|0)\right]\;
\left[1 - w_{u}(H_{2k}\!\uparrow\!H)\right] - 1.
\label{eq36}
\end{equation}
 
In order to solve this system of recursive equations we need the
formula for {\em the starting branch} of the relative
magnetization. Since {\em the negative saturation}
has been chosen as the initial state of the system it follows
from Eq. (\ref{eq9}) that if $H_{l} \Rightarrow H_{0} =
-\infty$, then
\[ w_{u}(H_{l}\!\uparrow\!H) \Rightarrow F_{u}(H|{\cal C}),  \]
and so the equation for the starting branch will be
\begin{equation}
m_{0}^{(u)}(-\infty\!\uparrow\!H|0) = 2\;F_{u}(H|{\cal C}) - 1.
\label{eq37}
\end{equation}
This branch can be also called {\em the limiting ascending branch}
because there is no branch below it.
If {\em the positive saturation} would be the initial state then
it is easy to show that {\em the limiting descending branch} can
be written in the form
\begin{equation}
m_{0}^{(d)}(\infty\!\downarrow\!H|0) = 2\;F_{d}(H|{\cal C}) - 1,
\label{eq38}
\end{equation}
where $F_{d}(H|{\cal C})$ is defined by the expression
(\ref{eq8}), and at the same time it is obvious that the limiting
descending branch has the property that there is no other
branch above it. The two limiting curves form {\em the major
hysteresis loop} which defines an area where all other loops
should be located.
 
By using the expression (\ref{eq37}) for the starting branch and
Eq. (\ref{eq35}) we can obtain the first descending
branch
\begin{equation}
m_{1}^{(d)}(H_{1}\!\uparrow\!H|0) = 2 F_{u}(H_{1}|{\cal C})\; [1 -
w_{d}(H_{1}\!\uparrow\!H)] - 1,
\label{eg39}
\end{equation}
which is attached to the limiting ascending branch at the point
$H_{1}$. This descending branch is called by Mayergoyz
\cite{mayergoyz91} {\em the first-order transition curve}. The
field $H_{1}$ where the first-order transition curve starts
from, will be called {\em start field.} Denote by $H_{2}$
the next reversal point  where the magnetizing
field begins again to increase. The corresponding ascending
branch, i. e., { \em the second-order transition curve} is given
by the formula:
\begin{equation}
m_{2}^{(u)}(H_{2}\!\uparrow\!H|0) = 2 w_{u}(H_{2}\!\uparrow\!H) +
[1 + m_{1}^{(d)}(H_{1}\!\downarrow\!H_{2})]\;
[1 - w_{u}(H_{2}\!\uparrow\!H)] - 1.
\label{eq40}
\end{equation}
This procedure can be continued and it is seen that there is
no need to take into account any special requirement
in order to describe the field dependence of the
average magnetization since the Markov points of the magnetizing
field determine exaclty the stochastic behaviour of the process.
 
\subsection{Stationarity of hysteresis loops}
 
Let us investigate now the variation of the magnetization for a
{\em special sequences of reversal fields}.
Let as suppose
that $H_{2k+1} = H_{u},\;\; \forall \;\; k=0,1,\ldots,$
while $H_{2k} = H_{d}, \;\; \forall \; k=1,2,\ldots,$ and
$H_{u} \geq H_{d}$, i.e. the magnetizing field is varying
between two extreme values $H_{u}$ and $H_{d}$. The field
variation which starts with a decrease of the the
external magnetic field $H$ from the reversal point $H_{u}$
until it reaches the next reversal point $H_{d}$ and then turns
to increase to the nearest $H_{u}$ value, is called {\em
magnetizing cycle}. The magnetizing cycle results in a hysteresis
loop called {\em minor hysteresis loop}. The {\em first cycle}
corresponds to the variation of the external
field between the reversal points
$H_{1} \Rightarrow H_{2} \Rightarrow H_{3}$, where
$H_{1} = H_{3} = H_{u}$ and $H_{2} = H_{d}$, while {\em the $k$-th
cycle} is done by the variation of the magnetizing
field between the reversal points $H_{2k-1} \Rightarrow H_{2k}
\Rightarrow H_{2k+1}$, where $H_{2k-1} = H_{2k+1} = H_{u}$
and $H_{2k} = H_{d}$ for $k=1,2,\ldots,\;\;$.
For {\em the descending branch of the $k$-th minor loop}
one can obtain from Eq. (\ref{eq35}) the following
expression:
\begin{equation}
m_{2k-1}^{(d)}(H_{u}\!\downarrow\!H|0) =
[1 + m_{2k-2}^{(u)}(H_{d}\!\uparrow\!H_{u}|0)]\;[1 -
w_{d}(H_{u}\! \downarrow\!H)] - 1.
\label{eq41}
\end{equation}
while for {\em the ascending branch of the $k$-th minor loop}
the relation
\begin{equation}
m_{2k}^{(u)}(H_{d}\!\uparrow\!H|0) =
2\;w_{u}(H_{d}\!\uparrow\!H) +
[1 + m_{2k-1}^{(d)}(H_{u}\!\downarrow\!H_{d}|0)]\;[1 -
w_{u}(H_{d}\! \uparrow\!H)] - 1
\label{eq42}
\end{equation}
can be derived from Eq. (\ref{eq36}).
By using Eq. (\ref{eq37}) we have
\begin{equation}
m_{1}^{(d)}(H_{u}\!\downarrow\!H|0) =
2\;F_{u}(H_{u}|{\cal C})\;[1 - w_{d}(H_{u}\!\downarrow\!H)] - 1,
\label{eq43}
\end{equation}
for the descending branch of the first minor loop and
\begin{equation}
m_{2}^{(u)}(H_{d}\!\uparrow\!H|0) =
2\;F_{u}(H_{u}|{\cal C})\;[1 -
w_{d}(H_{u}\!\downarrow\!H_{d})]\;[1 - w_{u}(H_{d}\!\uparrow\!H)] +
2\;w_{u}(H_{d}\!\uparrow\!H) - 1
\label{eq44}
\end{equation}
for the ascending branch of the same loop.
It is to note that according to the Mayergoyz's terminology the
first minor loop consists of a first-order descending
and a second-order ascending transition curves.
Following the reversal points of the magnetizing field this
procedure can be continued  and  we can obtain both the
descending and ascending branches of relative
magnetization for any minor loop.
 
One can prove {\em a very important limit theorem}, namely, there
exist two {\em limit curves}
\begin{equation}
\lim_{k\rightarrow \infty} m_{2k-1}^{(d)}(H_{u}\!\downarrow\!H|0)
= m_{d}(H_{u}\!\downarrow\!H),
\label{eq45}
\end{equation}
and
\begin{equation}
\lim_{k\rightarrow \infty} m_{2k}^{(u)}(H_{d}\!\uparrow\!H|0)
= m_{u}(H_{u}\!\downarrow\!H),
\label{eq46}
\end{equation}
which are determining a closed minor loop. In other words, the
magnetizing process becomes
stationary with increasing number of cycles. It means the system
"forgets" gradually its initial state by repeating the
magnetizing cycle. This forgetting process can be related to the
well-known  {\em accommodation process}. The original Preisach
model results in an {\em immediate formation} of the minor
hysteresis loop after only one cycle of back-and-forth
variation of the input between any two consecutive extremum
values. However, this consequence of the Preisach model
contradicts to many experimental facts
\cite{melville50}$^,$ \cite{carter50}
indicating the accommodation plays an important role
in magnetizing processes.  In order to describe the accommodation
process the traditional Preisach model was modified rather
artificially in the "moving" and the "product" models
\cite{torre87}. In contrast to these {\em the stochastic model
contains the phenomenon of accommodation inherently} and so there
is no need for any additional improving  the model.
 
The formulas for the limit curves defined by Eqs. (\ref{eq45}) and
(\ref{eq46}) can be obtained by some elementary calculations in
following forms:
\begin{equation}
m_{d}(H_{u}\!\downarrow\!H) = 2\;Q(H_{d}, H_{u})\;
w_{u}(H_{d}\!\uparrow\!H_{u})\;[1 - w_{d}(H_{u}\!\downarrow\!H)] -
1,
\label{eq47}
\end{equation}
and
\begin{equation}
m_{u}(H_{d}\!\uparrow\!H) = 1 - 2\;Q(H_{d}, H_{u})\;
w_{d}(H_{u}\!\downarrow\!H_{d})\;[1 - w_{u}(H_{d}\!\uparrow\!H)],
\label{eq48}
\end{equation}
where
\[ Q(H_{d}, H_{u}) = \left[w_{d}(H_{u}\!\downarrow\!H_{d}) +
w_{u}(H_{d}\!\uparrow\!H_{u}) -
w_{d}(H_{u}\!\downarrow\!H_{d})\;w_{u}(H_{d}\!\uparrow\!H_{u})
\right]^{-1}. \]
From these Eqs. two important relations can be derived, namely
\[ m_{d}(H_{u}\!\downarrow\!H_{u}) =
m_{u}(H_{d}\!\uparrow\!H_{u}),\;\;\;\; \mbox{and} \;\;\;\;
m_{d}(H_{u}\!\downarrow\!H_{d}) = m_{u}(H_{d}\!\uparrow\!H_{d}),
\]
which show that the return-point memory property is fulfiled for
the accommodated minor hysteresis loops. It is also obvious, that
the accommodated minor loops due to the same pair of reversal
fields $H_{d}$ and $H_{u} \geq H_{d}$ are not only congruent but
identical since the field values $H_{d}$ and $H_{u}$ unambiguously
determine the the branches of stationary loops. One has to mention
that the accommodated branches $m_{d}(\infty\!\downarrow\!H)$ and
$m_{u}(-\infty\!\uparrow\!H)$ are exactly identical with the
limiting descending and ascending branches which indicates the
consistency of the theory.
 
The explicit form of the expressions
$m_{d}(H_{u}\!\downarrow\!H)$ and
$m_{u}(H_{d}\!\uparrow\!H)$ which describe the descending and the
ascending branches of the stationary minor loop between two
reversal fields $H_{d}$ and $H_{u} \geq H_{d}$ has a great
advantage in numerical calculations in comparision with the
well-known Everett integral. It is to be noted that the
expressions (\ref{eq47}) and (\ref{eq48}) are suitable to describe
not only symmetrical but {\em asymmetrical hysteresis loops} too
and it is easy to show that symmetrical hysteresis loops
can be obtained only if the function $h(x, y)$ has a {\em mirror
symmetry} expressed by Eq.
\begin{equation}
h(x, y) = h(-y, -x).
\label{eq49}
\end{equation}
In the following the mirror symmetry of $h(x, y)$ will be
assumed.
 
It is worthwhile to derive the formula for {\em the
virgin curve of the magnetization} depending on the parameters of
the density function $h(x,y|{\cal C})$. After some simple
manipulations we obtain
\begin{equation}
m_{0}(H) = 2\;\frac{s_{1}(H)}{s_{1}(H) + s_{2}(H) -
s_{1}(H)\;s_{2}(H)} - 1,
\label{eq50}
\end{equation}
where
\begin{equation}
s_{1}(H) = \frac{\int_{-H}^{+H} dx\;\int_{-\infty}^{x} h(x, y) \;
dy}{\int_{-H}^{+\infty} dx\;\int_{-\infty}^{x} h(x, y) \;dy},
\label{eq51}
\end{equation}
and
\begin{equation}
s_{2}(H) = \frac{\int_{-H}^{+H} dy\;\int_{y}^{+\infty} h(x, y)
\;dx}{\int_{-\infty}^{+H} dy\;\int_{y}^{+\infty} h(x, y) \;dx.}
\label{eq52}
\end{equation}
If the function $h(x, y)$ satisfies the symmetry relation
(\ref{eq49}) then it is easy to prove that
\[ \lim_{H\rightarrow 0} m_{0}(H) = 0 \]
and {\em the initial susceptibility} defined by Eq.
\begin{equation}
\chi_{a} = \lim_{H \rightarrow 0} \frac{dm_{0}(H)}{dH} =
\frac{\int_{0}^{+\infty} h(x,0)\;dx}{\int_{0}^{+\infty}dx\;
\int_{-\infty}^{x} h(x, y)\;dy}
\label{eq53}
\end{equation}
is different from zero in contrary to the classical Preisach model
which gives a nonrealistic zero slope of the virgin curve at
$H=0$.
 
\section{Numerical calculations and discussion}
 
In order to compute the magnetization vs. field curves we have to
know the joint density function $h(x, y|{\cal C})$ of the U- and
D-fields.
Since these fields are the sum of many small random components it
is reasonable to assume that the central limit theorem is
approximately valid and so the function $h(x,y)$ in
$h(x, y|{\cal C})$ can be chosen in the form
\[ h(x, y) = \frac{1}{2\pi\;\sigma^{2}\;\sqrt{1 -
C_{r}^{2}}}\times \]
\begin{equation}
\times\exp\left\{-\frac{1}{2\sigma^{2}\;(1 - C_{r}^{2})}\;
\left[(x-H_{c})^{2} - 2\;C_{r}\;(x-H_{c})\;(y+H_{c}) +
(y+H_{c})^{2}\right]\right\},
\label{eq54}
\end{equation}
where the meaning of the constants $H_{c},\;\; \sigma$ and
$C_{r}$ is clear from the elements of the probability theory.
Figure \ref{fig3} shows the contour plot of $h(x, y|{\cal C})$
defined
by (\ref{eq6}) for the parameters $H_{c} = 0.2, \;\; \sigma = 0.6$
and $C_{r} = 0.5$. The contours are belonging to the
following values
of $h(x, y|{\cal C}) = 0.1, 0.2; 0.3(0.05)0.65$ and $0.682$. The
last one is slightly smaller than $\max_{(x,y)}\; h(x, y|{\cal
C}) = 0.682923...$.
The discontinouity along the line $y-x=0$ can be clearly seen
in the figure.
In the sequel this formula will be used in all of our numerical
calculations provided that the correlation coefficient $C_{r}$
is equal to zero, i. e. $h(x, y) = f(x) f(-y)$ where $f$ is
the density function of the normal distribution. This case
corresponds to {\em the product model} introduced by G. Biorci and
D. Pescetti \cite{biorci58} and used consequently by G. K\'ad\'ar
\cite{kadar87}$^{,}$  \cite{kadar89}$^{,}$ \cite{torre87}.
 
The relative magnetization vs. field curves are shown in Fig.
\ref{fig4}.  The parameter values used for the calculation
are $H_{c}=0.4, \;\; \sigma=0.6$ and $C_{r}=0$.
The curves {\bf LA} and {\bf LD}
correspond to the {\em limiting ascending and descending
branches}, while the curves indexed in the figure by
{\bf 1, 2, 3, 4} are {\em the first-,
second-, third- and fourth-order transition curves} defined
by the reversal points $H_{1}=1.2,\; H_{2}=-0.8,\; H_{3}=0.6,\;
H_{4}=-0.6$. It is worth noting that the all information
about the past history of the magnetizing process is transfered
by the state of the system in the last reversal point. For
example, the fourth-order transition curve {\bf 4} which is
plotted in the field interval $[-0.6, 1.5]$, is determined by the
state in the reversal point $H_{4}=-0.6$.
 
It is well-known that in the traditional Preisach model the minor
loops which describe the cyclic change of the magnetization
with back-and-forth variation of the magnetizing field
between the same two limiting values are congruent
and the formation of a closed minor loop is realized in
one cycle, i. e., {\em the accommodation process} is absent.
In contrast to this the stochastic model contains inherently
the accommodation process which is clearly demonstrated in
Fig. \ref{fig5}. For the sake of orientation
the limiting ascending branch {\bf LA} is also plotted in
Fig. \ref{fig5} where it is seen that
the descending branch of the first minor loop
starts from the point {\bf A} due to the first reversal field
$H_{1}=0.8$ and after reaching the
reversal point $H_{2}=H_{d}=-0.2$ it turns to increase to the
point {\bf B} which corresponds to the next reversal field
$H_{3}=H_{u}=0.8$. One can observe
that {\em the first minor loop} is not closed,
the point {\bf B} where the decreasing branch of {\em the
second minor loop} starts from,
occupies a higher position than the point {\bf A}, and
the end point {\bf C} of the increasing branch of the
second minor loop is found above  the point
{\bf B} but the distance between the points {\bf C} and {\bf B}
is smaller than that between the points {\bf B} and {\bf A}.
By repeating the magnetizing cycle between the
reversal fields $H_{d}=-0.2$ and $H_{u}=0.8$ the difference
between the branches of the same type becomes gradually
negligible, i.e., the branches converge to {\em limit curves}
which form finally a {\em closed stationary hysteresis loop}
denoted by {\bf LC}. The magnetizing curves measured by Carter and
Richards \cite{carter50} on silicon steel ($4.3\% Si$) are
surprisingly similar to that plotted in Fig. \ref{fig5}.
 
In order to demonstrate the speed of the convergence, the
non-accommodated relative magnetizations have been calculated in
the reversal point $H_{u}=0.8$ for the subsequent cycles.
Figure \ref{fig6} shows that the stationary (i. e., the
limit) value of the magnetization can be very well approached by
repeating the cycle 8-9 times in the case of
parameter values $H_{c}=0.4, \sigma=0.6$ and $H_{u}=0.3$.
 
The non-accommodated minor loops due to
the same pair of reversal fields are evidently
not congruent and generally are not closed. However, this
non-congruency has nothing in common with that introduced and
discussed in details by K\'ad\'ar \cite{kadar87}$^{,}$
\cite{gkadar87}. The non-congruency of the non-accommodated minor
loops bounded by the same field limits has a quite different
origin in the stochastic model, namely, the non-equilibrium
response of the system for the cyclic back-and-forth variation of
the external magnetic field between two consecutive reversal
points. It is obvious consequence of the non-stationarity of minor
loops that the return-point memory property is absent in these
loops.
 
In order to study the properties of non-congruency of this
type the first minor loops
belonging to different start fields are calculated.
Denote by $\Delta H = H_{u} - H_{d}$ the difference between the
consecutive reversal fields. For the characterization of the
non-accommodated first minor loops due to different start fields
$H_{1}$ let us introduce two parameters defined by
\[ {\bf W} = W(H_{1}, \Delta H) = \max_{H_{1}-\Delta H
\leq H \leq H_{1}} [m_{1}(H_{1}\!\downarrow\!H) -
m_{2}(H_{1}-\Delta H\!\uparrow\!H)], \]
and
\[ {\bf O} = O(H_{1}, \Delta H) =
m_{2}(H_{1}-\Delta H\!\uparrow\!H_{1}) -
m_{1}(H_{1}\!\downarrow\!H_{1}). \]
The dependence of these parameters on the start field $H_{1}$
is shown in Fig. \ref{fig7} for the parameter
values: $H_{d}=-0.2$ \and $\Delta H = 1$.
The author of the present paper is far not convinced whether the
experimental data contradict or support the non-congruency of this
type because of the lack of careful measurements.
 
It seems to be useful to investigate the remanence properties
of systems described by the stochastic model.
In Fig. \ref{fig8} the first-order descending curves which
start from different points of the ascending limiting branch
{\bf LA} can be seen. The curves starting from the points due
to the field values $H_{1}=1.6, H_{2}=1.4, H_{3}=1.2, H_{4}=1$
are plotted to {\em the points of remanences} {\bf R1, R2, R3,
R4}  which are obviously
different from the stationary (i. e., the accommodated) values.
The non-accommodated {\bf NR} and stationary remanences
{\bf SR} versus start field are shown in Fig. \ref{fig9}.
As it is seen the non-accommodated remanences can be negative
below a critical start field {\bf CR}
since the initial negative saturation has a significant
effect on the first-order transition curves. The stationary
remanence curve {\bf SR} calculated from the equations
(\ref{eq47}) and (\ref{eq48}) is non-negative in all points of
the start field interval.
 
The {\em influence of the parameter $\sigma$} on the shape of the
major hysteresis loop can be seen in Fig. \ref{fig10}. As it is
expected the larger is the parameter $\sigma$ the wider is the
hysteresis loop, i.e. the larger non-homogeneity in a system (e.g.
in a magnetic sample) results in a higher "coercive force".
 
It seems to be useful to calculate the {\em accommodated
(stationary) hysteresis loops} for different pairs of reversal
points $H_{d}$ and $H_{u} \geq H_{d}$. The hysteresis loops
plotted in Fig. \ref{fig11} correspond to the reversal points:
$H_{d}=-1.5,\; H_{u}=1.5$ (loop {\bf ML1}),
$H_{d}=-1,\; H_{u}=1$ (loop {\bf ML2}),
$H_{d}=-0.5,\; H_{u}=0.5$ (loop {\bf ML3}). For the
calculation we used the parameter values: $H_{c}=0.4$ and
$\sigma=0.6$. For the sake of completeness {\em the virgin curve}
{\bf VC} calculated by (\ref{eq50}) and the major loop
{\bf LL} bounded by the limiting ascending and descending
curves are also shown in the figure. The stochastic model
clearly shows that all accommodated minor loops corresponding to
cyclic inputs between the same two consecutive extremum values
are not only congruent but simply identical.
 
In Fig. \ref{fig12} three accommodated first-order minor loops
denoted by {\bf 1, 2, 3} can be seen. The descending branches of
the loops are started from the field values $H = 0.5, 0.3, 0$,
and each of the ascending branches returns exactly to the same
point that the corresponding descending branch left. The returning
curves have an apparent slope discontinuity with regard to the
major loop {\bf ALA}.
 
At this point it is worth to make a remark of somewhat
historical nature. As it is well-known  Preisach's idea
for his model was originated from the {\em quadratic Rayleigh
relation} which can be easily obtained \cite{becker39} assuming
a uniform distribution of the $U$- and $D$-fields over the
"Preisach triangle". It is interesting to note that in the
stochastic model the calculated hysteresis loops almost
perfectly coincide with that calculated by the Rayleigh formula
when the reversal fields $H_{d}$ and $H_{u} \geq H_{d}$ and so
the magnetizing field $H \in [H_{d}, H_{u}]$ are sufficiently
small. The hysteresis loop {\bf R} defined by reversal
points  $H_{u}=0.5$ and $H_{d}=-0.5$ in Fig. \ref{fig13} can be
very well approximated by the equations
\begin{eqnarray}
m_{a}^{d}(0.5\!\downarrow\!H) & = & C_{0}^{(d)} + C_{1}^{(d)}\;H +
C_{2}^{(d)}\;H^{2}, \nonumber \\
m_{a}^{u}(0.5\!\uparrow\!H) & = & C_{0}^{(u)} + C_{1}^{(u)}\;H +
C_{2}^{(u)}\;H^{2}, \nonumber
\end{eqnarray}
where
\begin{eqnarray}
C_{0}^{(d)} & = & - C_{0}^{(u)} = 0.13035..., \nonumber \\
C_{1}^{(d)} & = & C_{1}^{(u)} = 0.79301..., \nonumber \\
C_{2}^{(d)} & = & - C_{2}^{(u)} = - 0.54251... \;\;\; \nonumber
\end{eqnarray}
in the case of parameter values $H_{c}=0.2$ and $\sigma=0.6$.
In Fig. \ref{fig13} the squares \fbox{ } correspond to the values
calculated by the quadratic equations. The excellent aggreement
with the curves of the stochastic model indicates that the
Rayleigh law can be reproduced in a straightforward way in
the stochastic model.
 
This model differs from the original Preisach model in a very
essential point in relation to the {\em reversal point
susceptibility.} Namely, the non-zero initial susceptibility at
the turning points is an inherent property of the stochastic
model, while the traditional Preisach model can produce positive
initial slope only if the Preisach function is supposed to have
a Dirac-delta like singularity along the boundery of the
Preisach triangle, and that is a rather artificial requirement
introduced by Mayergoyz \cite{mayergoyz91}. The
susceptibility vs. magnetizing field is seen in Fig. \ref{fig14}
for the parameter values: $H_{c} = 0.2$ and $\sigma = 0.6$. The
shape of the calculated curve can be expected on the basis of
physical considerations and  corresponds to those found
experimentally.
 
The estimation of the joint density function $h(x, y|{\cal C})$
from measured hysteresis curves was beyond the scope of our
present theoretical consideration. Of course, one may attempt
in simple cases to estimate the parameters of a plausible
density function (e. g., (\ref{eq54})) by an appropriate data
evaluation procedure.
 
\section{Conclusions}
It has been shown that the Preisach model of hysteresis can be
replaced by a new model based on  exact concepts  of the
probability theory. In this model the phenomenon of hysteresis
has been described as a {\em stochastic process} defined on a set
of all possible values of the control parameter the reversal
(turning) points of which are Markov points of the process.
The one dimensional distribution function of the stochastic
process has been exactly determined and the magnetizations
versus up and down magnetic fields have been calculated as
expectation values of the stochastic process.
It has been proven that the magnetizing process becomes stationary
with increasing number of magnetizing cycles. It means that for
the description of the accommodation process there
is no need of any artificial auxiliary assumption since the
stochastic model contains the phenomenon of accommodation
inherently. In general case the model is able to describe
the symmetric as well as the asymmetric hysteresis. In
relatively small magnetizing fields the
quadratic Rayleigh law  can be easily obtained from the
equations of the stochastic model. It is important to note that
the turning point susceptibilities have non-zero finite values
in contrary to the traditional Preisach model which does not take
consequently into account the random nature of the
elementary switching process. Finally, the stochastic model shows
that all stationary loops corresponding to the same two
limiting values of the magnetizing field are equivalent but the
non-stationary loops are  non-congruent and in general not closed.
 
\section*{Acknowledgments}
The author wish to acknowledge the
stimulating discussions with  Mr. G. K\'ad\'ar who helped very
much in completing the present paper.

\begin{figure}
\caption{A possible realization of the transition
${\cal S}^{(+)} \Leftrightarrow {\cal S}^{(-)}$.
\label{fig1}}
\end{figure}
 
\begin{figure}
\caption{Two equivalent $H(t)$ curves.
\label{fig2}}
\end{figure}
 
\begin{figure}
\caption{Contour plot of $h(x, y|{\cal C})$ defined by Eq. $(6)$
where $h(x,y)$ is given by Eq. $(57)$ with parameter values:
$H_{c} = 0.2, \sigma = 0.6$ and $C_{r} = 0.5$.
\label{fig3}}
\end{figure}
 
\begin{figure}
\caption{Limiting branches {\bf LA, LD} and
magnetization versus field curves starting from the reversal
points $H_{1}=1.2$, $H_{2}=-0.8$, $H_{3}=0.7$ and
$H_{4}=-0.6$. The magnetization curves are
indicated by {\bf 1, 2, 3} and {\bf 4}.
\label{fig4}}
\end{figure}
 
\begin{figure}
\caption{Accommodation process of the minor loop in consecutive
magnetizing cycles between the reversal points $H_{d}=-0.2$ and
$H_{u}=0.8$. The loop {\bf LC} is the accommodated minor
loop.
\label{fig5}}
\end{figure}
 
\begin{figure}
\caption{Convergence of the relative magnetization in the reversal
point $H_{u}=0.8$ with increasing number of cycles to
the limit (i. e., the stationary) value.
\label{fig6}}
\end{figure}
 
\begin{figure}
\caption{Width {\bf W} of the first-order minor loops and the
difference {\bf O} between the values of the descending and
ascending branches in the reversal point $H_{u}=0.8$ versus start
field.
\label{fig7}}
\end{figure}
 
\begin{figure}
\caption{The limiting ascending branch {\bf LA} and four
first-order descending curves ending in non-accommodated
remanences denoted by {\bf R1, R2, R3} and {\bf R4}.
\label{fig8}}
\end{figure}
 
\begin{figure}
\caption{The non-accommodated {\bf NR} and the stationary
{\bf SR} relative remanences versus start field due to
different points of the ascending limiting branch.
\label{fig9}}
\end{figure}
 
\begin{figure}
\caption{Influence of the parameter $\sigma$ on the shape of the
major hysteresis loop in the case of $H_{c}=0.4$.
\label{fig10}}
\end{figure}
 
\begin{figure}
\caption{The virgin curve {\bf VC}, the major loop {\bf LL}
and three accommodated hysteresis loops
{\bf ML1, ML2, ML3} calculated for different pairs of reversal
points in the case of parameter values $H_{c}=0.4$ and
$\sigma=0.6$.
\label{fig11}}
\end{figure}
 
\begin{figure}
\caption{Three accommodated first-order minor loops denoted by
{\bf 1, 2, 3} and the shifted ascending branch {\bf ALA}.
\label{fig12}}
\end{figure}
 
\begin{figure}
\caption{The hysteresis loop between "small" reversal points and
the quadratic Rayleigh curves denoted by $\Box$.
\label{fig13}}
\end{figure}
 
\begin{figure}
\caption{The irreversible susceptibility versus magnetizing field
$H$.
\label{fig14}}
\end{figure}

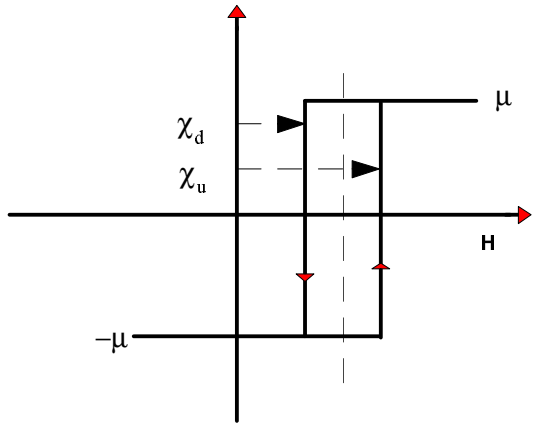
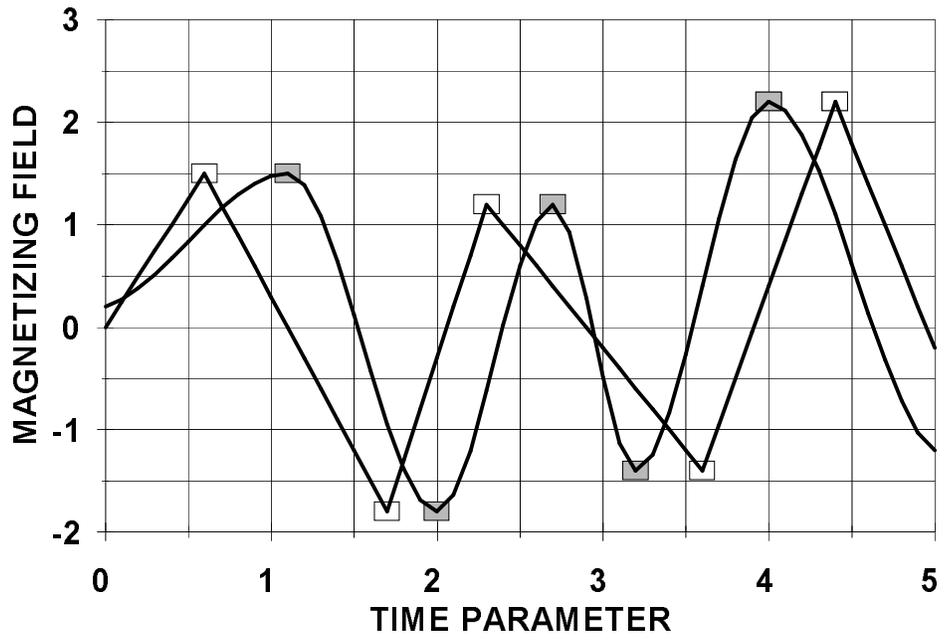
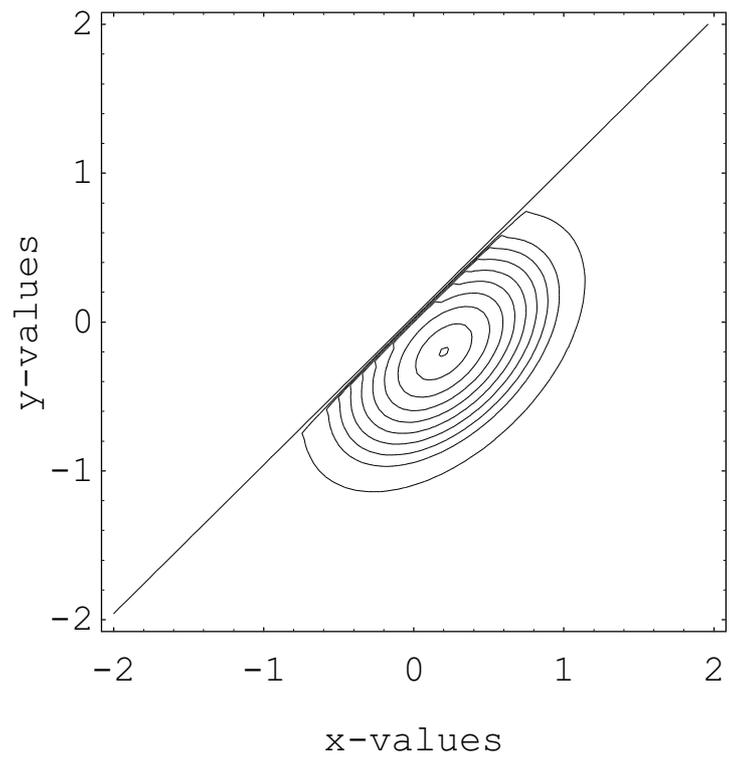
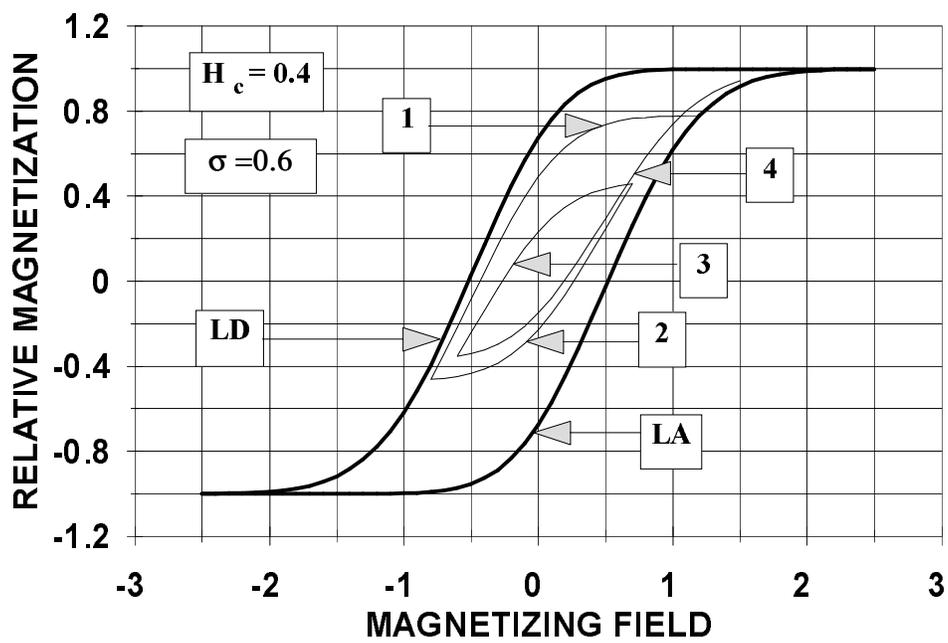
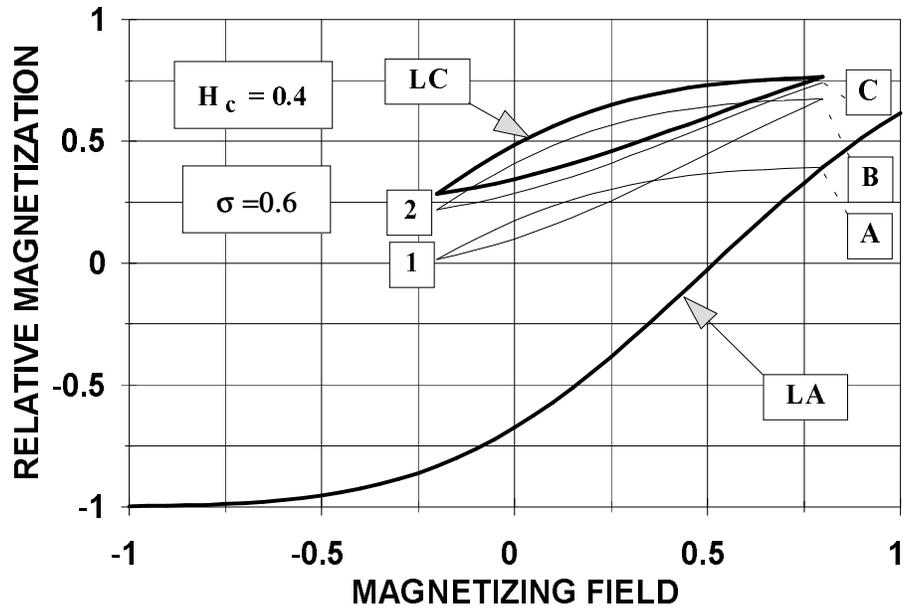
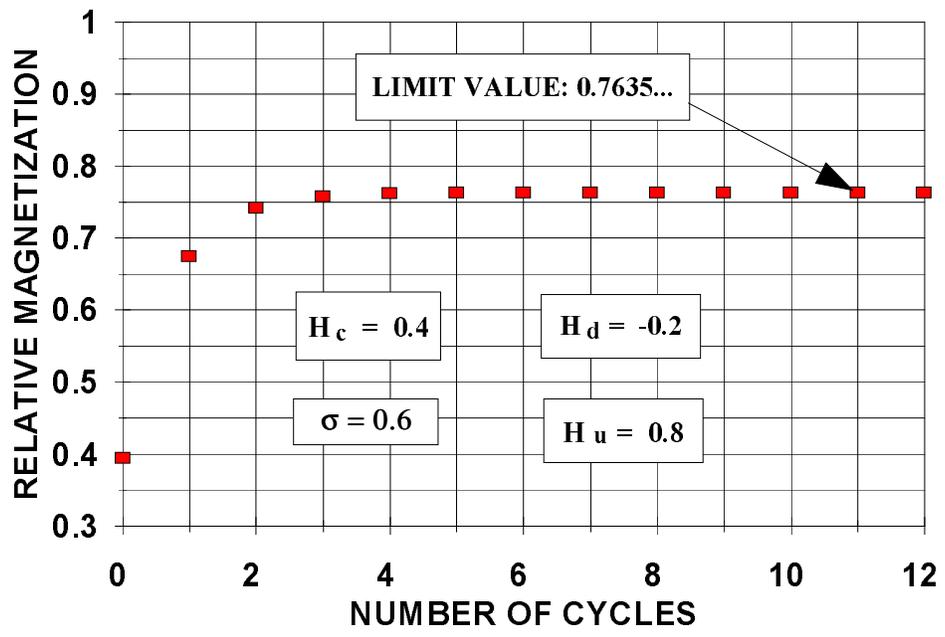
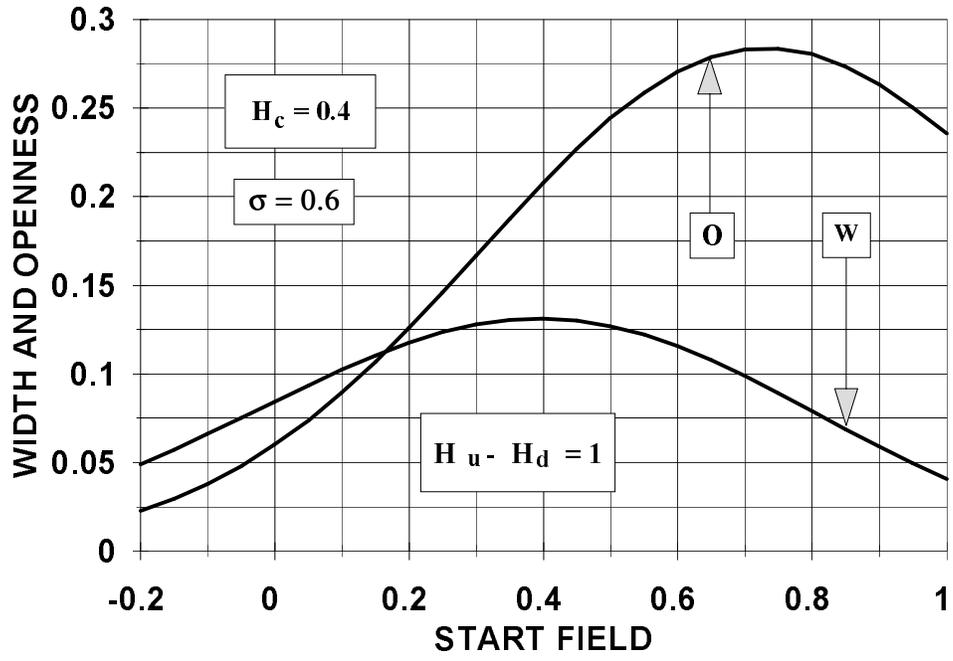
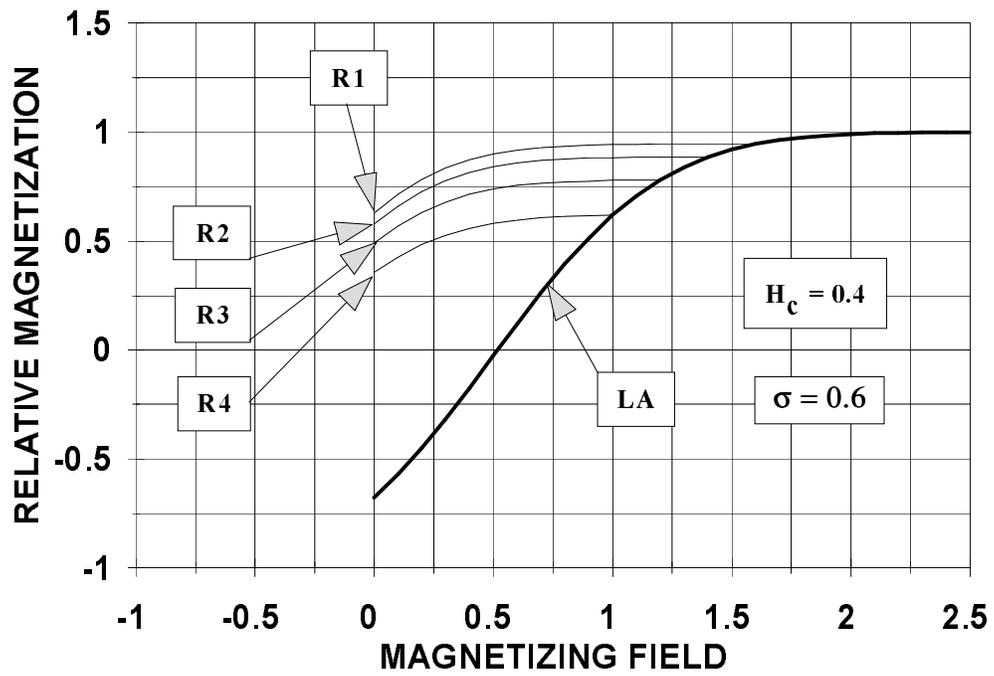
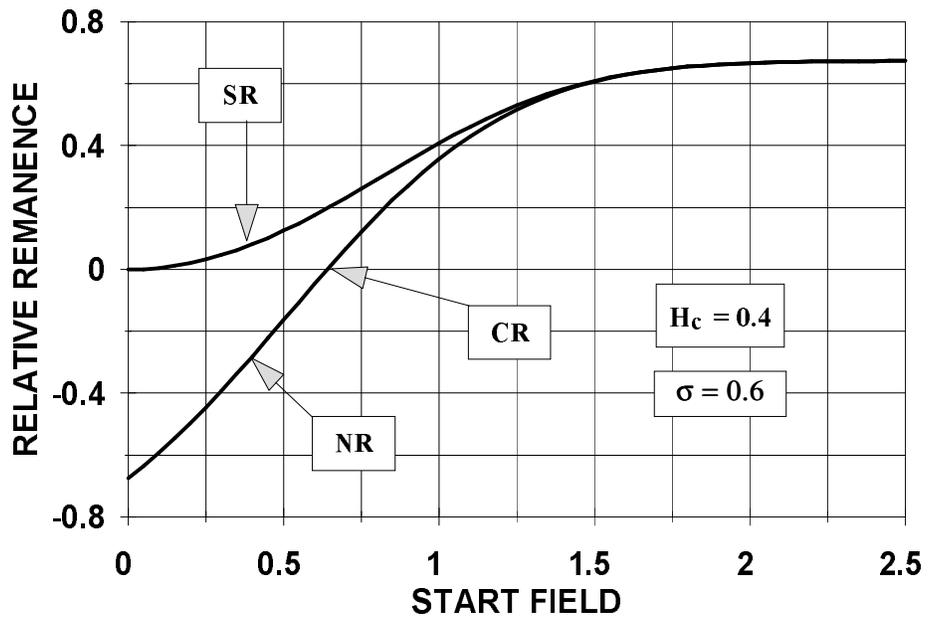
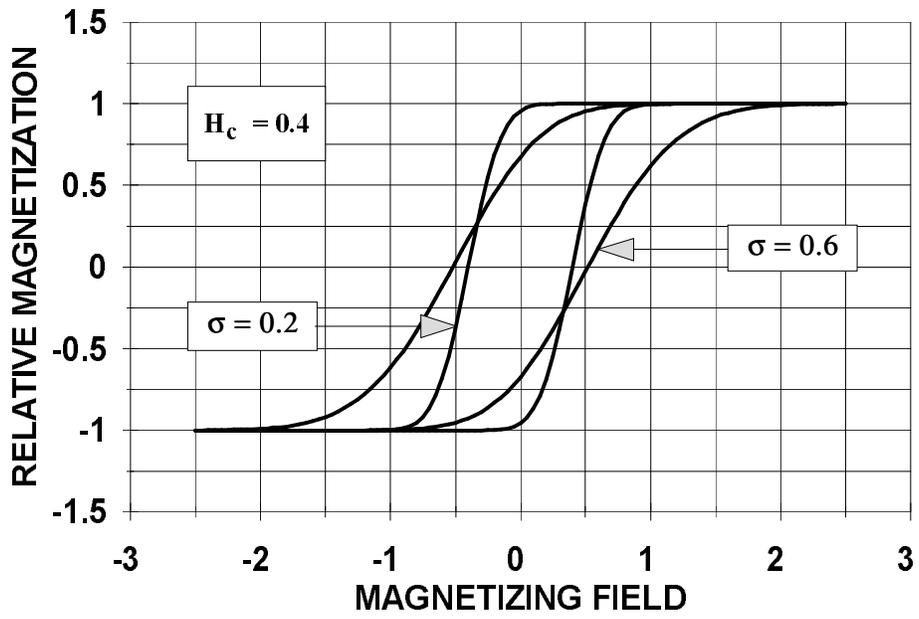
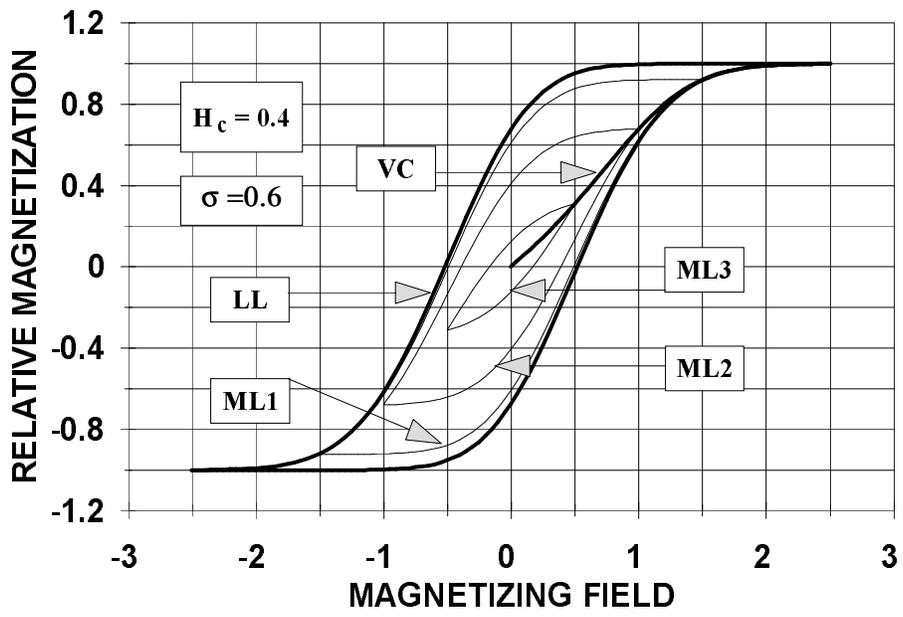
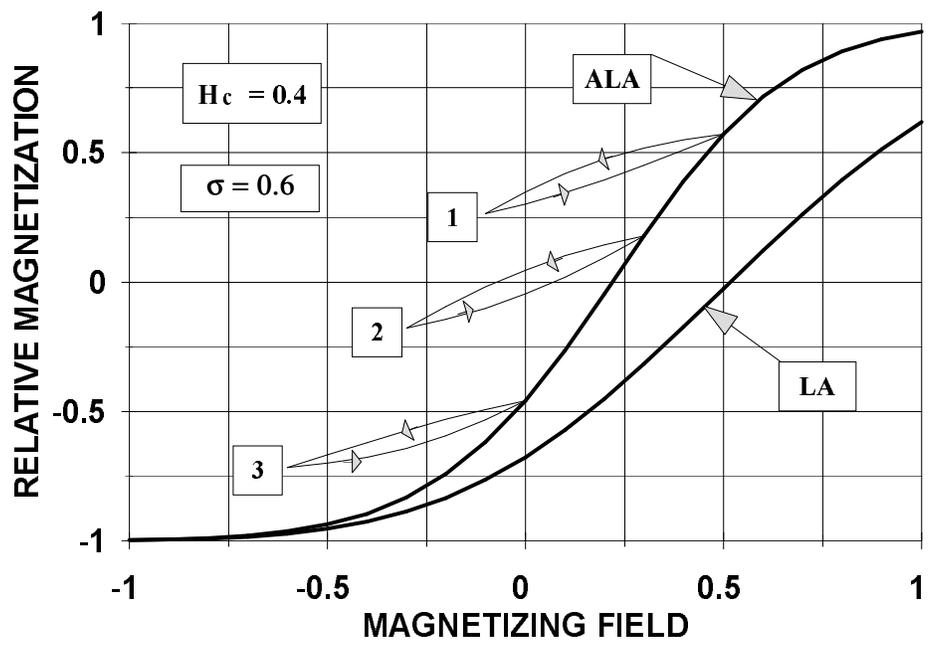
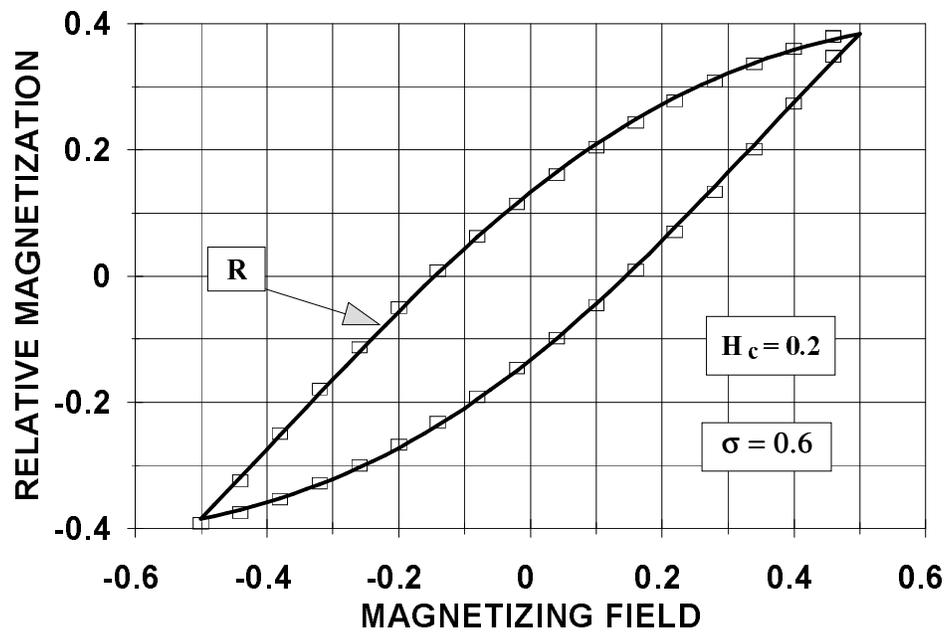
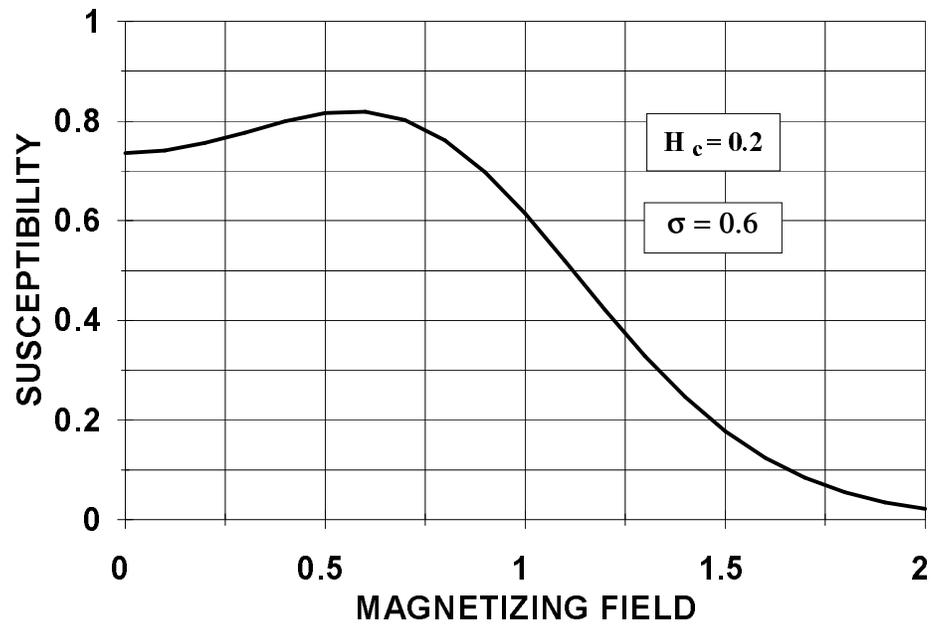
\end{document}